\documentclass[10pt,conference]{IEEEtran}
\usepackage{amsmath,amssymb,amsfonts,bm} 
\usepackage{algorithmic}
\usepackage{algorithm}
\usepackage{array}
\usepackage{caption}
\usepackage{subcaption}
\usepackage{textcomp}
\usepackage{stfloats}
\usepackage{bigints}
\usepackage{url}
\usepackage{verbatim}
\usepackage{graphicx}
\usepackage{mathtools}
\usepackage{dsfont}
\usepackage{esvect}
\usepackage{accents}
\usepackage{cite}
\usepackage{pbalance}
\usepackage{xcolor}
\newcolumntype{M}[1]{>{\centering\arraybackslash}m{#1}}
\newcommand*{\QEDA}{\null\nobreak\hfill\ensuremath{\blacksquare}}%

\newcommand\myeqa{\mathrel{\stackrel{\makebox[0pt]{\mbox{\normalfont\scriptsize (a)}}}{=}}}
\newcommand\myeqb{\mathrel{\stackrel{\makebox[0pt]{\mbox{\normalfont\scriptsize (b)}}}{=}}}
\newcommand\myeqc{\mathrel{\stackrel{\makebox[0pt]{\mbox{\normalfont\scriptsize (c)}}}{=}}}

\begin{document}

% paper title
% Titles are generally capitalized except for words such as a, an, and, as,
% at, but, by, for, in, nor, of, on, or, the, to and up, which are usually
% not capitalized unless they are the first or last word of the title.
% Linebreaks \\ can be used within to get better formatting as desired.
% Do not put math or special symbols in the title.
\title{Communication and Sensing Coverage of UAV Corridor under Beam Misalignment Error}
%  
%
% author names and IEEE memberships
% note positions of commas and nonbreaking spaces ( ~ ) LaTeX will not   brea k
% a structure at a ~ so this keeps an author's name from being broken across
% two lines.
% use \thanks{} to gain access to the first footnote area
% a separate \thanks must be used for each paragraph as LaTeX2e's \thanks
% was not built to handle multiple paragraphs
%
\author{Harris K.~Armeniakos$^*$, 
        Petros S. Bithas$^{**}$,
        Konstantinos Maliatsos$^*$,
        Christos Masouros\IEEEauthorrefmark{2},
         \\and Athanasios G.~Kanatas$^*$   \\
       $^*$Department of Digital Systems, University of Piraeus, Piraeus, Greece \\
       $^{**}$Department of Digital Industry Technologies, National and Kapodistrian University of Athens, Greece \\
        \IEEEauthorrefmark{2}Department of Electronic and Electrical Engineering, University College London, WC1E 7JE London, U.K\\
        Email: \{harmen,kanatas\}@unipi.gr, pbithas@dind.uoa.gr, kmaliat@aegean.gr, c.masouros@ucl.ac.uk
        % <-this % stops a space
%\thanks{The authors are with the Department
%of Digital Systems, University of Piraeus, 18534 Piraeus, Greece (e-mail:
%\{harmen,kanatas\}@unipi.gr).}% <-this % stops a space
}
\maketitle

% As a general rule, do not put math, special symbols or citations
% in the abstract or keywords.
\begin{abstract}
In this paper, a unmanned aerial vehicle (UAV) corridor-assisted network is proposed to study the joint sensing and communication (JCAS) coverage probability (CP) of a terrestrial (BS) equipped with a realistic three-dimensional (3D) 3GPP antenna pattern. By employing the 3D binomial point process (BPP) to model the spatial locations of the UAV-UEs, a fixed number of UAV-UEs are deployed in an aerial corridor modeled as a finite cylinder. The BS exploits radar sensing capabilities to track a known UAV-UE under a 3D beam misalignment error, which follows a multivariate normal distribution. Subsequently, under clutter and uplink communication interference, exact-form analytical expressions for JCAS CP are derived and used to analyze the performance at the terrestrial BS. The results highlight corridor-length saturation at high UAV heights and the beamwidth trade-off between interference suppression and misalignment sensitivity.
\end{abstract}

% Note that keywords are not normally used for peerreview papers.
\begin{IEEEkeywords}
Aerial highways, beam misalignment, joint communication and sensing (JCAS), UAV corridors.  
\end{IEEEkeywords}

% For peer review papers, you can put extra information on the cover
% page as needed:
% \ifCLASSOPTIONpeerreview
% \begin{center} \bfseries EDICS Category: 3-BBND \end{center}
% \fi
%
% For peerreview papers, this IEEEtran command inserts a page break and
% creates the second title. It will be ignored for other modes.
\IEEEpeerreviewmaketitle

\section{Introduction}
As a new economic form, low-altitude economy exploits low-altitude airspace (usually within 1000 meters above
the ground) to carry out various missions, creating commercial and social values. The flying activities are conducted by various UAVs, enabling a variety of applications, including transportation,
agriculture, and disaster monitoring \cite{10955337}. In UAV-to-ground communications, UAVs elevated position increases the probability of LoS propagation, but also introduces severe interference challenges with terrestrial networks \cite{Cai2021JSAC}. However, as UAV traffic becomes denser, unconstrained 3D mobility is increasingly difficult to manage from both communication and air-traffic perspectives. This has motivated the concept of aerial highways or UAVs corridors, where UAVs fly within predefined routes or volumes to reduce collision risk, simplify traffic management, protect conventional air traffic, and concentrate wireless coverage in specific aerial regions. Such corridors have been studied in the context of cellular-network optimization through antenna-tilt, transmit-power, and beamwidth design \cite{KarimiBidhendi2024UAVCorridors,Benzaghta2025UAVCorridors}, enhancement of corridor coverage, and stochastic-geometry-based analysis in finite corridor-assisted networks \cite{Armeniakos2024LCOMM,Armeniakos2026TCOMM}. Therefore, aerial corridors constitute not only an air-traffic-management mechanism, but also a communication-oriented deployment paradigm for reliable UAV networks.

Building on this corridor-assisted perspective, integrated sensing and communication (ISAC), also referred to as JCAS, is a natural technology candidate for future aerial networks. JCAS enables sensing and communication functionalities to share spectrum, hardware, waveforms, and signal-processing resources, and is therefore considered a key 6G technology \cite{Armeniakos2026COMST}. This integration is particularly attractive for UAV-enabled networks, where size, weight, power, and payload constraints favor joint sensing--communication designs, while UAV mobility and LoS-dominant links support coverage, localization, tracking, and beam alignment \cite{Jing2024ISACSky,Pang2024DynamicISAC,Zhang2024MultiUAVISAC,Bithas2025OJVT,Mei2024DelayISAC,Cheng2025NetworkedISAC}. In \cite{Jing2024ISACSky}, UAV trajectory and bandwidth allocation were optimized under energy constraints, while in \cite{Pang2024DynamicISAC}, UAV's reflected signals  were exploited to track vehicle states and dynamically adapt ISAC beams. In \cite{Zhang2024MultiUAVISAC}, UAV trajectory, user association, and beamforming were jointly optimized, while in \cite{Bithas2025OJVT}, the performance of sensing-assisted UAV-assisted communications was analytically studied. In \cite{Mei2024DelayISAC}, delay violation in clustered ISAC UAV networks was analyzed under transient antenna misalignment, while \cite{Cheng2025NetworkedISAC} considered networked ISAC through the joint design of coordinated beamforming and UAV trajectories.

In aerial JCAS systems, antenna misalignment represents a critical impairment, since UAV and mmWave/THz links often rely on narrow directional beams to compensate for path loss and enable accurate sensing. In practice, UAV mobility, wind-induced platform instability, blockage, imperfect sensing estimates, and delayed beam switching can shift the target away from the main beam, resulting in loss of coverage, outage, and delay degradation \cite{Chen2022BeamAlignment,Guo2025Beamwidth,Mei2024DelayMisalignment,Xue2024BeamManagement,Huang2025ISACBeamManagement}. In \cite{Chen2022BeamAlignment}, SSB-based sensing was used to reduce blockage- and mobility-induced beam misalignment in THz/mmWave networks. In \cite{Guo2025Beamwidth}, sensing-assisted mmWave coverage was analyzed under angular estimation errors, and the beamwidth was optimized to balance beamforming gain and misalignment probability. In \cite{Mei2024DelayMisalignment}, transient antenna misalignment in ISAC UAV networks was investigated from a delay-violation perspective. Related studies have also emphasized low-latency beam management under mobility \cite{Xue2024BeamManagement} and ISAC-enabled beam management with sensing errors in THz networks \cite{Huang2025ISACBeamManagement}. However, these works typically rely on simplified sectored beams, abstract gain-loss models, or non-standard antenna assumptions. 

\emph{Contributions:} Motivated by the above, this paper develops a stochastic-geometry framework for analyzing the performance of a UAV corridor-assisted aerial network. The UAV-UEs are confined within an aerial corridor modeled as a finite 3D cylinder, which provides a more realistic representation of structured UAV mobility than line-segment corridor models, while also accounting for probabilistic LoS/NLoS propagation between the terrestrial BS and the UAV-UEs. Furthermore, instead of adopting idealized sectored antennas, terrestrial BS is assumed to be equipped with a standardized 3D 3GPP antenna pattern, so that off-boresight radiation and angular-dependent gain degradation are explicitly captured. Practical beam pointing errors are incorporated through a 3D beam-misalignment model, where the elevation and azimuth errors follow a correlated multivariate normal distribution. Based on this setup, the JCAS CP is characterized under communication and clutter interference, thus quantifying the joint impact of corridor geometry, blockage, realistic antenna directivity, and spatial beam misalignment on aerial JCAS performance.

\section{System Model}
\subsection{Spatial and Network Modeling}
Consider a finite UAV-assisted network that integrates UAV-UEs and a terrestrial BS. The UAV-UEs are assumed to hover in an aerial corridor, with rules enforced by the appropriate traffic authorities, modeled as a finite \emph{3D cylinder}. In particular, the cylinder has been proposed as the most reasonable geometric model to approximate a real UAV corridor \cite{yu2025hybrid}. Consequently, we assume that the UAV-UEs are independently and uniformly distributed within the finite cylindrical region $\mathcal{C}(L,R)$, where $\mathcal{C}(L,R)$ represents a 3D cylinder centered at $\mathbf{o'}=[0,0,h]^T$, with axial extent from $-L/2$ to $L/2$, having a radius $R$, and being located $h$ meters above the ground. Without loss of generality, the on-ground projection of the center of the UAV corridor, namely $\mathbf{o'}$, crosses the origin of the coordinates $\mathbf{o} = [0,0,0]$. Subsequently, it is assumed that a predefined and fixed number of $N$ UAV-UEs equipped with isotropic antennas are distributed on $\mathcal{C}(L,R)$ and their spatial locations form a 3D BPP $\Phi$. It should be noted that the proposed framework can be applied directly to the practical case with directional antennas for UAV-UEs. A  terrestrial BS equipped with a 3D 3GPP-based antenna pattern is assumed to be centered at the origin $\mathbf{o}$, as shown in Fig. 1. Let $\{\mathbf{w}_i\} \equiv \Phi$ denote the spatial locations of the UAV-UEs. Consequently, $\{\mathbf{w}_i\}=\{(x_i,y_i,z_i)\}_{i=1:N}$ denotes the set of 3D spatial locations in Cartesian coordinates of the UAV-UEs in the corridor, while $\|\mathbf{w}_i\|$ denotes the Euclidean distance between the $i$-th UAV-UE and the terrestrial BS. In the proposed UAV-ISAC scheme, it is assumed that different time-frequency resources are used for sensing and communications functionalities \cite{khalili2024efficient}. The terrestrial BS enables radar sensing capabilities to track a UAV-UE's movement trajectory, while the tracked UAV-UE performs an uplink communication with the terrestrial BS, as also shown in Fig. 1.

\begin{figure}[!t]
    \centering
    \includegraphics[keepaspectratio,width=3.5in]{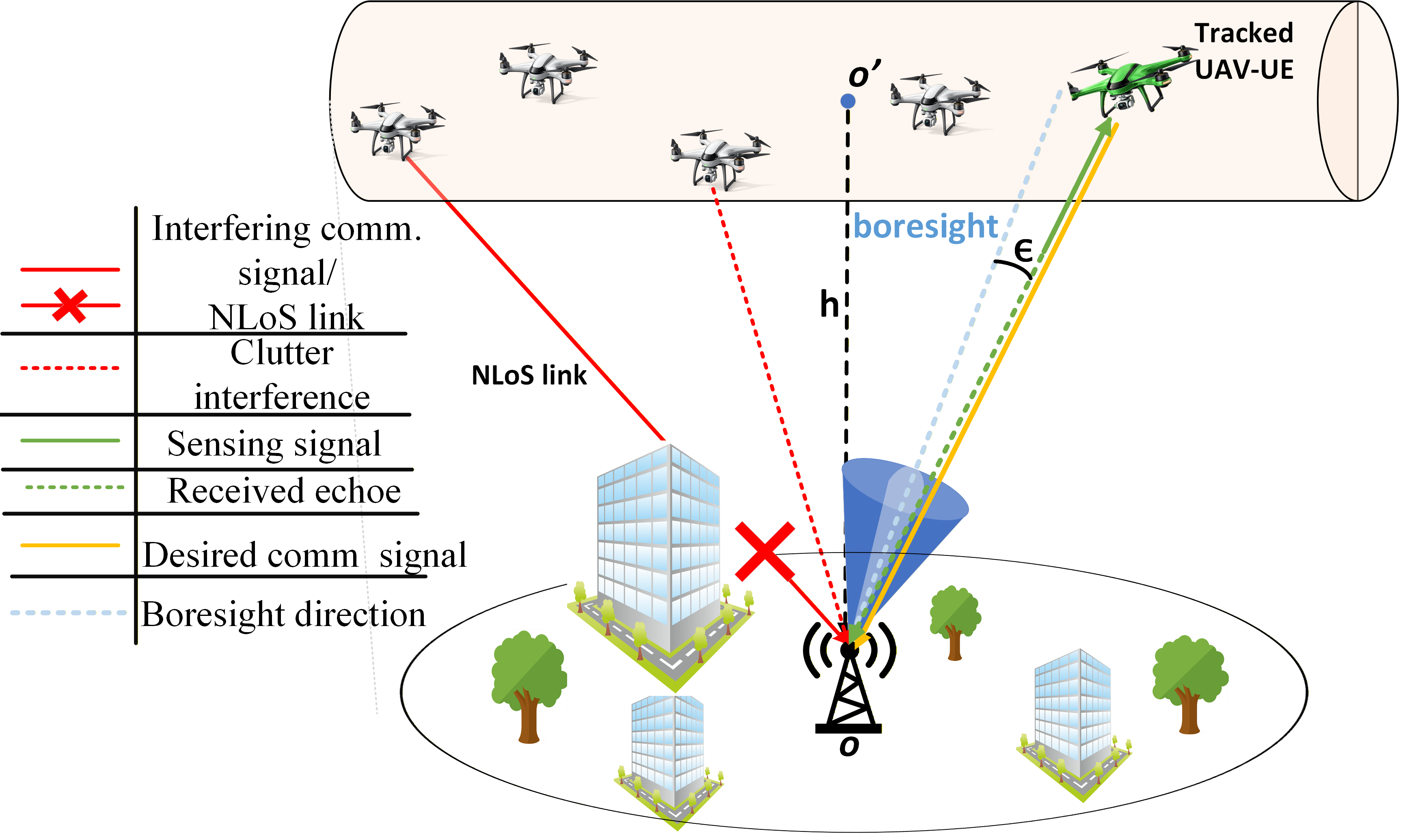}
       \caption{ Illustration of the UAV corridor-assisted JCAS network with a terrestrial BS equipped with a 3D 3GPP antenna pattern.}
    \label{fig:Fig2}
\end{figure}

\subsection{Antenna and Misalignment Error Modeling}
The terrestrial BS is equipped with beamforming capabilities to support JCAS operation with the target UAV-UE. The BS antenna array forms a directive beam towards the target UAV-UE to support both sensing and communication functionalities. In this work, a realistic 3GPP-based antenna pattern is adopted, so that antenna gain is modeled as an angle-dependent quantity rather than as a constant value within the $3$~dB beamwidth. When directed towards $(\theta_0,\phi_0)$, this antenna pattern has  vertical and  horizontal radiation power patterns (in dB) given by
\begin{align}\label{eq6}
       &G_{3gpp}^V(\theta,\theta_0) = -{\rm{min}}\Bigg\{ 12\Bigg(\frac{\theta-\theta_0}{\theta_{3dB}} \Bigg)^2, SLA \Bigg\}, \\ \nonumber
       &G_{3gpp}^H(\phi,\phi_0) = -{\rm{min}}\Bigg\{ 12\Bigg(\frac{\phi-\phi_0}{\phi_{3dB}} \Bigg)^2, SLA \Bigg\},   
 \end{align}
and a 3D antenna pattern $G_{3gpp}^{3D}(\theta,\theta_0,\phi,\phi_0)= 
       G_{{\rm{max}}}-{\rm{min}}\big\{-G_{3gpp}^V(\theta,\theta_0) -G_{3gpp}^H(\phi,\phi_0),SLA\big\}$, where $\theta_{3dB}$ and $\phi_{3dB}$ are the vertical and horizontal $3$~dB beamwidths and $SLA = 30$~dB is the front-to-back ratio. The multiplicative antenna gain factor is indicated by $g_{3gpp}^{3D}(\theta,\theta_0,\phi,\phi_0)$ and the multiplicative
maximum antenna gain factor of the maximum antenna gain $G_{{\rm{max}}}$ is indicated by $g_{{\rm{max}}}$. The terrestrial BS is assumed to steer its beam towards the tracked UAV-UE.

% The terrestrial BS takes advantage of the sensing data gathered from the tracked UAV-UE to infer its instantaneous position/trajectory and updates the beam steering direction accordingly.

However, under practical JCAS operation scenarios, beam misalignment between the BS and the tracked UAV-UE may occur, and the boresight direction of the BS's antenna fails to point directly towards the exact location of the tracked UAV-UE. The impact of beam misalignment is captured by the 3D beamsteering errors, as shown in Fig. 1. In this work, triggered by \cite{11030947}, it is assumed that the 3D misalignment error $\mathbf{\epsilon} = \begin{bmatrix} \theta_{mis} \\ \phi_{mis} \end{bmatrix} $ follows a zero mean \emph{multivariate Gaussian distribution}, i.e., $\mathbf{\epsilon} \sim \mathcal{N}(\mathbf{0},\mathbf{\Sigma}_{mis})$ with PDF denoted by $f_{\boldsymbol{\epsilon}}(\mathbf{\epsilon})$. The covariance matrix $\mathbf{\Sigma}_{mis}$ is given by  $\mathbf{\Sigma}_{mis} = \begin{bmatrix}
\sigma_{\theta}^2 & \rho \sigma_{\theta} \sigma_{\phi} \\
\rho \sigma_{\theta} \sigma_{\phi} & \sigma_{\phi}^2 
\end{bmatrix}$,
where $\sigma_{\theta}^2, \sigma_{\phi}^2$ denotes the variance of the error in the elevation and azimuthal angle, respectively, and $\rho$ captures the correlation between the two angle errors. In this work, $\sigma_{\theta}, \sigma_{\phi}$ is defined as a fraction of the corresponding $3$~dB beamwidths and therefore $\sigma_{\theta} = c \cdot \theta_{3dB}$ and $\sigma_{\phi} = c \cdot \phi_{3dB}$, with $c$ denoting a scaling constant.

\subsection{Blockage and Channel Models}
A probabilistic LoS/NLoS blockage model is adopted to capture the blockage effect in UAV-assisted networks, where the LoS link probability is approximated by a simple modified Sigmoid function. Accordingly, the LoS link probability between the $i$-th UAV-UE and the terrestrial BS is given by $ P_L(\mathbf{w}_i) = \frac{1}{\alpha \exp \big[-b(\arctan\big(z_i/\sqrt{x_i^2+y_i^2}\big)\frac{180}{\pi}-\alpha)\big]}$,
where $(\alpha,b)$ denote the model's environment related parameters. In particular, they may capture various practical scenarios, e.g., Suburban (6.7309, 0.4434), Urban (9.6117, 0.1581) and Dense urban (12.0810, 0.1139). The NLoS link probability is $P_N(\mathbf{w}_i)=1-P_L(\mathbf{w}_i)$. The blockage effect leads to different path loss exponents, denoted $\alpha_L$ and $\alpha_N$, for LoS and NLoS links, respectively. The path loss law is then given by
\begin{equation}
l(\mathbf{w}_i)= \left\{
\begin{array}{ll}
     \Big(\sqrt{x_i^2 + y_i^2 + z_i^2}\Big)^{-\alpha_L} & {\rm{w.p.}} \quad P_L(\mathbf{w}_i) \\
      \Big(\sqrt{x_i^2 + y_i^2 + z_i^2} \Big)^{-\alpha_N}& {\rm{w.p.}} \quad P_N(\mathbf{w}_i)\\
\end{array}. 
\right. 
\end{equation}
In UAV-assisted networks, since LoS and NLoS links coexist, it is assumed that the signal amplitude follows the Nakagami distribution with the parameter $m_t(t \in \{L, N\})$. Therefore, the power gains of the channels $h_{i}$ follow the Gamma distribution, i.e, $h_{i} \sim $Gamma$\big(m_t,\frac{1}{m_t}\big)$ with the shape and scale parameters of $h_{i}$ being $m_t$ and $1/m_t$, respectively. Note that $\mathbb{E}[h_{i}]=m_t \frac{1}{m_t}=1$ and the values of $m_t$ are restricted to integers for analytical tractability.

\subsection{Association Policy and SIR Definition}
\subsubsection{UAV-UE Association Policy}
The terrestrial BS selects a specific UAV-UE for continuous tracking and fine trajectory adjustment along the UAV corridor. Let \(\mathbf{w}_0\) denote the location of the tracked UAV-UE.
The set of the remaining, non-tracked UAV-UEs is denoted by
\(\Phi^{!}=\Phi\setminus\{\mathbf{w}_0\}\), and is used to model clutter and uplink interference.

% In this work, the tracked UAV-UE  is chosen uniformly at random among the set of $N$ UAV-UEs distributed as a 3D BPP on  $\mathcal{C}$. In this work, the tracking functionality is captured over time through the different spatial realizations of the 3D BPP. Subsequently, the tracked UAV-UE performs uplink communication with the terrestrial BS.

\subsubsection{Communication and Sensing SIR Definition}
The terrestrial BS measures the received echo power reflected by the tracked UAV-UE, while the remaining \(N-1\) UAV-UEs in the corridor contribute clutter interference. For the rest of the analysis, the index $\{0\}$ denotes the tracked UAV-UE. Following the standard radar equation, the sensing SIR associated with the tracked UAV-UE in the terrestrial BS is expressed as
\begin{equation}
{\rm{SIR}}_s = \frac{P_{\rm{echo}}}{I_s} = \frac{g_{{\rm{mis}}}^{\boldsymbol{\epsilon}}(\theta, \theta_{0},\phi,\phi_{0})^2 h_{0}^s \sigma_0 l(\mathbf{w}_0)^2}{\sum\limits_{i \in \Phi^!} g_{{\rm{mis}}}^{\boldsymbol{\epsilon}}(\theta_i, \theta_{0},\phi_i,\phi_{0})^2   h_{i}^s \sigma_i  l(\mathbf{w}_i)^2}, 
\end{equation}
where $g_{{\rm{mis}}}^{\boldsymbol{\epsilon}}(\theta, \theta_{0},\phi,\phi_{0}) = g_{3gpp}^{3D}(\theta,\theta_{0}+\theta_{mis},\phi,\phi_{0}+\phi_{mis})$, while $\sigma_i$ denotes the random radar cross-section (RCS) \footnote{Typically, RCS is angle-dependent. Nevertheless, as shown in \cite{11159279}, the Swerling-based RCS models match almost perfectly with experimental data while on the same time do not further increase the analytical complexity.}. By adopting the Swerling I model, the RCS of any UAV-UE is modeled as an exponential RV with a common marginal PDF $f_{\sigma_i}(\sigma) = \frac{1}{\overline{\sigma}}\exp(-\sigma/\overline{\sigma})$, where $\overline{\sigma}$ denotes the average RCS. 

Meanwhile, the tracked UAV-UE is assumed to perform uplink communication with the BS, while the BS suffers from interference caused by the rest $(N-1)$ active UAV-UEs\footnote{In the communication phase, the interfering UAV-UEs are modeled by an independent BPP, denoted by \(\Phi^{!!}\), which is distinct from the BPP used to characterize the UAV-UEs involved in the sensing/tracking phase. In the more realistic case where the interferer locations across the sensing and communication phases are dependent, the corresponding aggregate interference terms become coupled, and the analysis would require conditioning on the entire BPP $\Phi$. This would result in a substantially different and much more involved mathematical treatment, which lies beyond the scope of this work.}. The latter captures a worst-case interference scenario. The communication SIR from the $i$-th tracked UAV-UE received at the  terrestrial BS's antenna array is given by
\begin{equation}
{\rm{SIR}}_c = \frac{S_0}{I_c}=\frac{g_{{\rm{mis}}}^{\boldsymbol{\epsilon}}(\theta_0, \theta_{0},\phi_0,\phi_{0}) h_{0}^c l(\mathbf{w}_0) }{\sum\limits_{i \in \Phi^{!!}} g_{{\rm{mis}}}^{\boldsymbol{\epsilon}}(\theta_i, \theta_{0},\phi_i,\phi_{0}) h_{i}^c l(\mathbf{w}_i)},
\end{equation}
where  $I_c$ denotes aggregate uplink interference from the
UAV-UEs to the  terrestrial BS.

\section{JCAS Performance Analysis}

\emph{Definition 1:} \emph{The JCAS CP $\mathcal{P}_{cs}(\theta_c, \theta_s)$, is defined as the probability that the received SIR of the communication and sensing functions at the BS are simultaneously not less than predefined thresholds $\theta_c$ and $\theta_s$, respectively, required for successful communication and tracking, i.e., $\mathcal{P}_{cs}(\theta_c, \theta_s)  \triangleq \mathbb{P}[ {\rm{SIR}}_{c}>\theta_c, {\rm{SIR}}_s>\theta_s]$. }

%Note that in the general case, $k^*_c, k^*_s$ can be different ports. It is worth mentioning that since communication and sensing functionality share the same spatial topology, the JCAS coverage probability cannot be obtained simply as $\mathcal{P}_{cs}(\theta_c, \theta_s) \neq \mathcal{P}_{c}(\theta_c) \mathcal{P}_{s}(\theta_s)$. Next, the sensing and communication coverage probability are first obtained conditioned on the spatial location of the BS and the tracked UAV-UE. 

\subsection{Sensing Coverage Probability}
The statistical characterization of both sensing and communication CP requires accounting for the Euclidean distances and angular coordinates of the UAV-UEs. Therefore, the joint PDF of the 3D spatial location of a UAV-UE in terms of the Cartesian coordinates is first derived as a key intermediate result in  CP analysis. 

\emph{Lemma 1:} \emph{The joint PDF $f_{\mathbf{w}_i}(x_i,y_i,z_i)$ is given by }
\begin{equation}
f_{\mathbf{w}_i}(x_i,y_i,z_i) = \frac{1}{|\mathcal{C}(L,R)|},
\end{equation}
\emph{where $|\mathcal{C}(L,R)|=\pi R^2L$ and $x_i \in[-L/2, L/2]$, $y_i \in\big[-\sqrt{R^2-(z-h)^2}, \sqrt{R^2-(z-h)^2}\big]$ and $z_i\in [h-R,h+R]$.}

\textit{Proof.} Since the UAV-UE locations form a uniform BPP in the finite cylindrical corridor \(\mathcal{C}(L,R)\), each UAV-UE location \(\mathbf{w}_i=(x_i,y_i,z_i)\) is independently and uniformly distributed over \(\mathcal{C}(L,R)\). Therefore, its joint spatial PDF is
\begin{equation} \label{eq_proof1}
    f_{\mathbf{w}_i}(x_i,y_i,z_i) = \frac{1}{|\mathcal{C}(L,R)|}, 
\end{equation}
where the corridor is defined as $\mathcal{C} = \big\{(x_i,y_i,z_i): x \in [-\frac{L}{2},\frac{L}{2}] ,\quad y^2+(z-h)^2\leq R^2\big\}$. By solving the inequality in \eqref{eq_proof1} for $y$ given a $z$ value, it yields $y^2 \leq  R^2 -(z-h)^2  \rightarrow |y| \leq \sqrt{R^2 -(z-h)^2}$ from which the final range of $y$ is obtained. Next, the restriction \(R^2-(z-h)^2 \geq 0\) gives
\(|z-h|\leq R\), from which the range of \(z\) follows as in Lemma~1. \QEDA

\emph{Lemma 2:} \emph{Conditioned on the location  $\mathbf{w}_0$ of the tracked UAV-UE and the misalignment error $\mathbf{\epsilon}$, the conditional Laplace transform  $\mathcal{L}_{I_s}(s|\mathbf{w}_0)$ of the aggregate clutter interference power distribution is given by}
\begin{align}
&\mathcal{L}_{I_s}(s|\mathbf{w}_0,\mathbf{\epsilon}) = \Big[  \int_{0}^{\infty} \int_{\mathcal{C}} \sum_{t \in  \{L,N\} } \frac{1}{\Big(1+\frac{s  \sigma_i g_{{\rm{mis}}}^{\boldsymbol{\epsilon}}(\mathbf{w}_i|\mathbf{w}_0)^2 }{m_t l^{-2}(\mathbf{w}_i)}\Big)^{m_t}} \\
& \times P_t(\mathbf{w}_i) f_{\mathbf{w}_i}(\mathbf{w}_i) f_{\sigma_i}(\sigma_i) {\rm{d}} \mathbf{w}_i {\rm{d}} \sigma_i\Big]^{N-1}, \nonumber 
\end{align} 
\emph{where $g_{{\rm{mis}}}^{\boldsymbol{\epsilon}}(\mathbf{w}_i|\mathbf{w}_0) = g_{{\rm{mis}}}^{\boldsymbol{\epsilon}}(\theta_i, \theta_{0},\phi_i,\phi_{0})$ and the angles in $g_{{\rm{mis}}}^{\boldsymbol{\epsilon}}(\theta_i, \theta_{0},\phi_i,\phi_{0})$  are given by $\phi_i=\rm{atan2}(y_i,x_i),\phi_0=\rm{atan2}(y_0,x_0), \theta_i = \tan^{-1}\big(\sqrt{x_i^2+y_i^2}/z_i\big),\theta_0 = \tan^{-1}\big(\sqrt{x_0^2+y_0^2}/z_0\big)$ and {\rm{atan2}} denotes the four quadrant arctan function \cite{press2007numerical}.} 

\textit{Proof.} See Appendix A. \QEDA

%Since the location $\mathbf{w}_0$, the LoS state and the misalignment error $\mathbf{\epsilon}$ is common for both sensing and communication, the conditional sensing CP is first derived.

\emph{Proposition 1:} \emph{The conditional sensing CP $\mathcal{P}_{s}(\theta_s|\mathbf{w}_0,\mathbf{\epsilon},t)$ of the  terrestrial BS is given by}
\begin{equation} 
\begin{split}
&\mathcal{P}_{s}(\theta_s|\mathbf{w}_0,\mathbf{\epsilon},t) =\int_{0}^{\infty} \sum_{k=0}^{m_t - 1}  \frac{{(-s_0(\mathbf{w}_0,\mathbf{\epsilon}))}^k }{ k!} \\
& \times{\Bigg[ \frac{\partial^{k} \mathcal{L}_{I_s}(s|\mathbf{w}_0,\mathbf{\epsilon})}{\partial{s^k}}\Bigg]}_{s=s_0(\mathbf{w}_0,\mathbf{\epsilon})} f_{\sigma_0}(\sigma_0) {\rm{d}}  \sigma_0, 
\end{split}
\end{equation}
\emph{for $s_0(\mathbf{w}_0,\mathbf{\epsilon}) = \frac{m_t \theta_s}{\sigma_0 g_{{\rm{mis}}}^{\boldsymbol{\epsilon}}(\theta_0,\phi_0)^2l^2(\mathbf{w}_0)}$, $t \in \{L,N\}$ and $\theta_0,\phi_0$ given by Lemma 2}.

\textit{Proof.} Due to page limitation, only the sketch of the proof is provided. Conditioned on $\mathbf{w}_0,\mathbf{\epsilon},c$, the proof for deriving Proposition 1 follows conceptual lines similar to those presented in \cite{TWC}. Finally, since the RCS of the target UAV-UE $\sigma_0$ is independent of the aggregate clutter interference, averaging over its PDF $f_{\sigma_0}(\sigma_0)$ yields Proposition 1.  \QEDA

\subsection{Communication Coverage Probability}
The tracked UAV-UE communicates with the terrestrial BS, and the $N-1$ UAV-UEs are assumed to interfere with the BS.

\emph{Lemma 3:} \emph{Conditioned on the location  $\mathbf{w}_0$ of the tracked UAV-UE and the misalignment error $\mathbf{\epsilon}$, the  Laplace transform  $\mathcal{L}_{I_c}(s|\mathbf{w}_0,\mathbf{\epsilon})$ of the aggregate interference power distribution is given by}
\begin{align}
&\mathcal{L}_{I_c}(s|\mathbf{w}_0,\mathbf{\epsilon}) = \Big[   \int_{\mathcal{C}} \sum_{t \in  \{L,N\} } \Big(1+\frac{s   g_{{\rm{mis}}}^{\boldsymbol{\epsilon}}(\mathbf{w}_i|\mathbf{w}_0)^2 }{m_t l^{-2}(\mathbf{w}_i)}\Big)^{-m_t} \\
& \times P_t(\mathbf{w}_i) f_{\mathbf{w}_i}(\mathbf{w}_i)  {\rm{d}} \mathbf{w}_i \Big]^{N-1}. \nonumber 
\end{align} 

\textit{Proof.} The proof for deriving Lemma 3 follows similar conceptual lines as the proof for deriving Lemma 2 and hence it is omitted here for brevity. \QEDA

%Having derived $\mathcal{L} _{I_c}(s|x_0,y_0)$, the conditional communication coverage probability at the BS conditioned on the spatial location $\mathbf{w}_i$ of the tracked UAV-UE and the location $\mathbf{p}_0$ of the BS can now be derived.

\emph{Proposition 2:} \emph{The conditional communication CP $\mathcal{P}_{c}(\theta_c|\mathbf{w}_0,\mathbf{\epsilon},t)$ of the  terrestrial BS is given by}
\begin{equation} 
\begin{split}
&\mathcal{P}_{c}(\theta_c|\mathbf{w}_0,\mathbf{\epsilon},t) = \sum_{k=0}^{m_t - 1}  \frac{{(-s_0(\mathbf{w}_0,\mathbf{\epsilon}))}^k }{ k!} \\
&\times {\Bigg[ \frac{\partial^{k} \mathcal{L}_{I_c}(s|\mathbf{w}_0,\mathbf{\epsilon})}{\partial{s^k}}\Bigg]}_{s=s_0(\mathbf{w}_0,\mathbf{\epsilon})} \, 
\end{split}
\end{equation}
\emph{for $s_0(\mathbf{w}_0,\mathbf{\epsilon}) = \frac{m_t \theta_c}{\ g_{{\rm{mis}}}^{\boldsymbol{\epsilon}}(\theta_0,\phi_0)^2l^2(\mathbf{w}_0)}$, $t \in \{L,N\}$ and $\theta_0,\phi_0$ given by Lemma 2}.

\textit{Proof.} The proof for deriving Proposition 2 follows similar conceptual lines as the proof for deriving Proposition 1 and hence it is omitted here for brevity. \QEDA

\subsection{JCAS Coverage Probability}
The JCAS CP can now be derived by jointly deconditioning the product of Proposition 1 and Proposition 2 as in the following Theorem. 

\emph{Theorem 1:} \emph{The JCAS CP of a terrestrial BS in a UAV corridor-assisted network is given by}
\begin{align} 
 &\mathcal{P}_{cs}(\theta_c, \theta_s) \\\nonumber 
 &=  \int_{\mathcal{C}}  \iint_{\boldsymbol{\epsilon}}  \sum_{t \in \{L,N\}} P_t(\mathbf{w}_0) \mathcal{P}_{c} (\theta_c|\mathbf{w}_0,\boldsymbol{\epsilon},t)  \mathcal{P}_{s}(\theta_s|\mathbf{w}_0,\boldsymbol{\epsilon},t) \\\nonumber
 &\times f_{\mathbf{w}_0}(\mathbf{w}_0)f_{\boldsymbol{\epsilon}}(\boldsymbol{\epsilon}) {\rm{d}} \mathbf{w}_0  {\rm{d}} \boldsymbol{\epsilon}. \nonumber
\end{align}

\textit{Proof.} See Appendix B. \QEDA

%\textit{Proof.} Recalling Definition 1, the JCAS coverage probability is given by
%\begin{equation} \label{proofTh3}
%\begin{split}
%&\mathcal{P}_{cs}(\theta_c, \theta_s) = \mathbb{E}[\mathds{1}({\rm{SIR}}_{k^*_c}^c \geq \theta_c) \mathds{1}({\rm{SIR}}_{k^*_s}^s \geq \theta_s)] \\
%&= \mathbb{E}_{\mathbf{p}_0,\mathbf{w}_i}[\mathds{1}({\rm{SIR}}_{k^*_c}^c \geq \theta_c | \mathbf{p}_0,\mathbf{w}_i) \times \mathds{1}({\rm{SIR}}_{k^*_s}^s \geq \theta_s |\mathbf{p}_0,\mathbf{w}_i)] \\
%&  \myeqa  \mathbb{E}_{\mathbf{p}_0,\mathbf{w}_i}[\mathbb{P}({\rm{SIR}}_{k^*_c}^c \geq \theta_c |\mathbf{p}_0,\mathbf{w}_i) \times \mathbb{P}({\rm{SIR}}_{k^*_s}^s \geq \theta_s |\mathbf{p}_0,\mathbf{w}_i)],
%\end{split}
%\end{equation}
%where (a) follows directly after applying the expectations which result to the conditional probabilities that have been proved in Theorem 1 and Theorem 2. After substituting Theorem 1 and Theorem 2 in \eqref{proofTh3} and applying the expectation over $\mathbf{p}_0$ w.r.t $f_{\mathbf{p}_0}(x_0,y_0)$  and $w_i$ w.r.t $f_{w_i}(w_i)$ , Theorem 3 yields. \QEDA

%\emph{Remark:} \emph{The results obtained in Theorem 1 and Theorem 2 can be exploited to directly investigate the marginal sensing and coverage probabilities, respectively for a predetermined spatial location of the BS. In a similar vein, the JCAS coverage probability obtained in Theorem 3 can directly be exploited to investigate the JCAS performance for a known location of the BS. This is particular insightful, if one is interested in investigating the JCAS performance towards the network edge.} 

\section{Numerical Results \& Discussions}
In this section, numerical results are presented to evaluate the JCAS performance achieved in the proposed  UAV corridor-assisted network. The accuracy of the analytical results is verified by comparing them with the simulated ones obtained using Monte Carlo simulations. The following parameters have been used, unless otherwise stated: $L = 100$~meters, $R=5$ meters, $\alpha_L = 2.1$, $\alpha_N = 4$, $m_L = 3$, $m_N = 1$,  $N = 10$, $h=100$ meters, $(\phi_{3dB},\theta_{3dB}) = (45^o,45^o)$, $G_{\rm{max}}=8$~dBi, $SLA=30$~dB and $\overline{\sigma} = 0.7 m^2$. 

\begin{figure}[!t]
    \centering
    \includegraphics[keepaspectratio,width=3.45in]{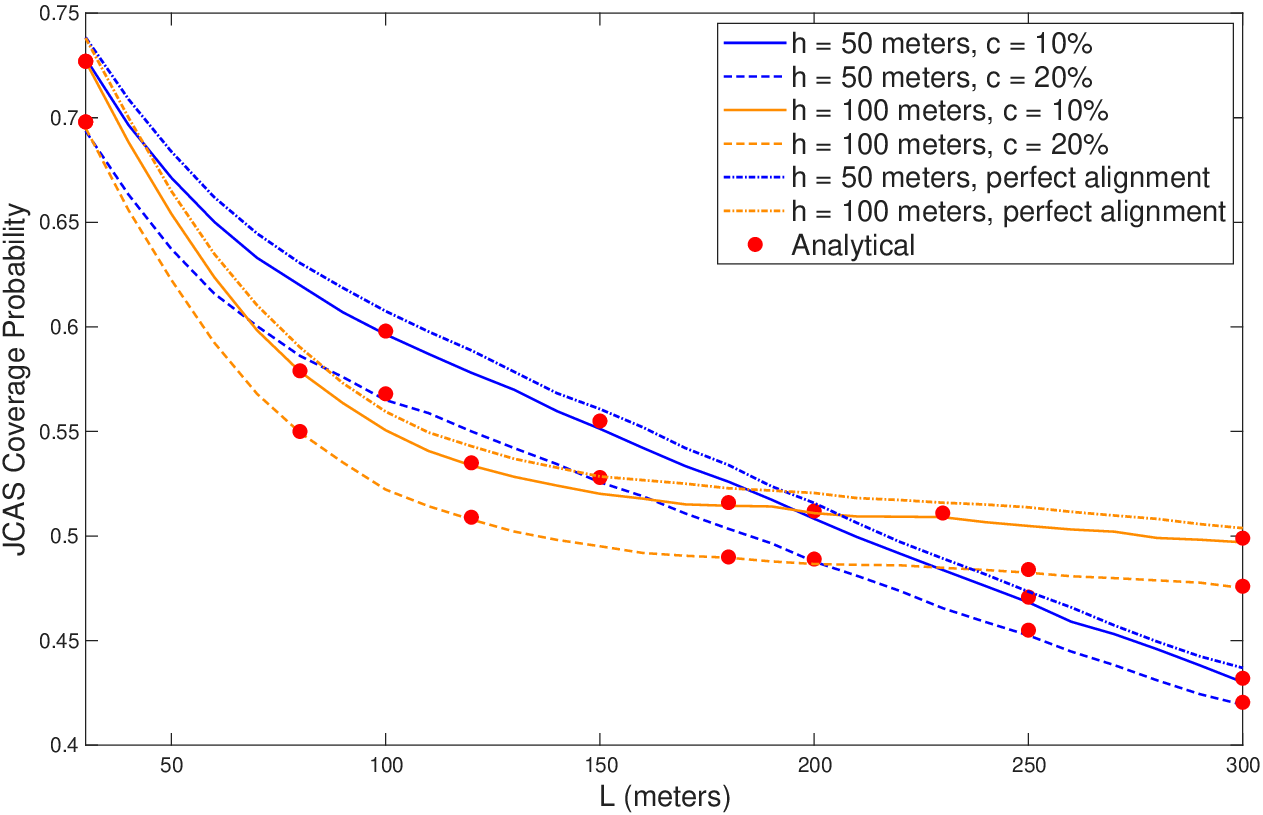}
       \caption{JCAS CP versus $L$ for different values of $h$ and misalignment errors.}
    \label{fig:Fig2}
\end{figure}

Fig. \ref{fig:Fig2} presents the  analytical (denoted by markers) and simulated JCAS CP, obtained by Theorem 1, versus $L$ for different values of $h$ and misalignment errors. A constant $c=10$ \% corresponds to a deviation of $4.5^{o}$ in the multivariate normal distribution, while a constant $c=20$ \% corresponds to a deviation of $9^{o}$. The JCAS CP under perfect beam alignment is also simulated for completeness. A first observation is that a deviation of $4.5^{o}$, causes a misalignment error that can be considered tolerant for both values of $h$, while for a deviation of $9^{o}$, the misalignment error causes a profound reduction in JCAS performance. A key observation is that the JCAS CP decreases with an increase of corridor length until it eventually stabilizes only for the higher deployment heights. Indeed, at sufficiently high altitude, the network becomes less sensitive to corridor-length expansion because the LoS advantage and the spread of interferers compensate for the additional pathloss.

\begin{figure}[!t]
    \centering
    \includegraphics[keepaspectratio,width=3.45in]{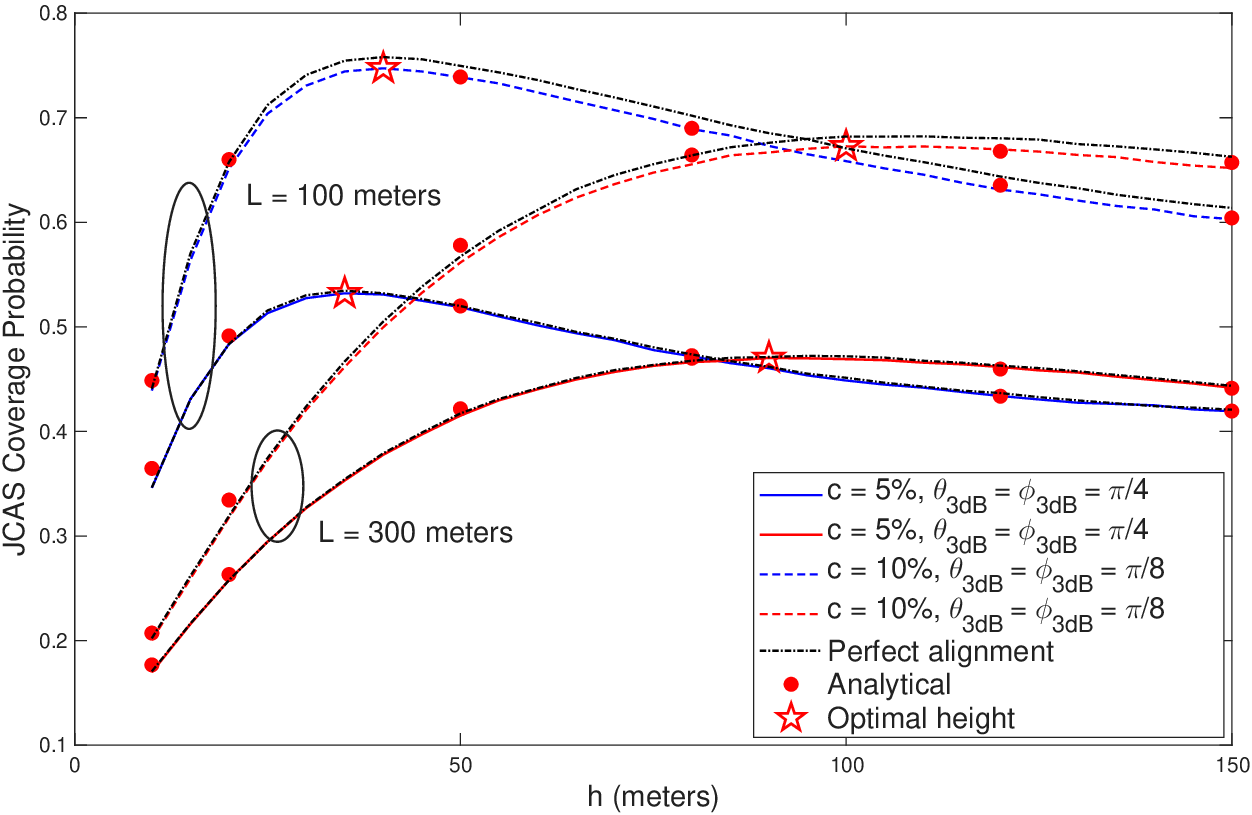}
       \caption{JCAS CP versus $h$ for different values of $L$, $3$~dB beamwidths, and misalignment errors.}
    \label{fig:Fig3}
\end{figure}

Fig. \ref{fig:Fig3} presents the  analytical (denoted by markers) JCAS CP, obtained by Theorem 1, versus $h$ for different values of $L$, 3dB beamwidths and misalignment errors. Both the solid and dashed curves share the same angular standard deviation, i.e., $\sigma_{\theta}=\sigma_{\phi} = 2.25^{o}$. A first observation is that for the wider beams, an angular deviation of up to $2.25^{o}$, hardly affects the JCAS performance due to beam misalignment error.  Subsequently, an interesting trade-off can be drawn: A narrower beam at the BS's antenna reduces interference more effectively, but it is also more sensitive to pointing errors. In contrast, a wider beam is more resistant to misalignment errors but more susceptible to interference. Finally, it is observed that there exists an optimal height of UAVs $h^*$ (denoted by a red star) to maximize the JCAS CP. In particular, this optimal height increases with the length of the corridor. 

\begin{figure}[!t]
    \centering
    \includegraphics[keepaspectratio,width=3.9in]{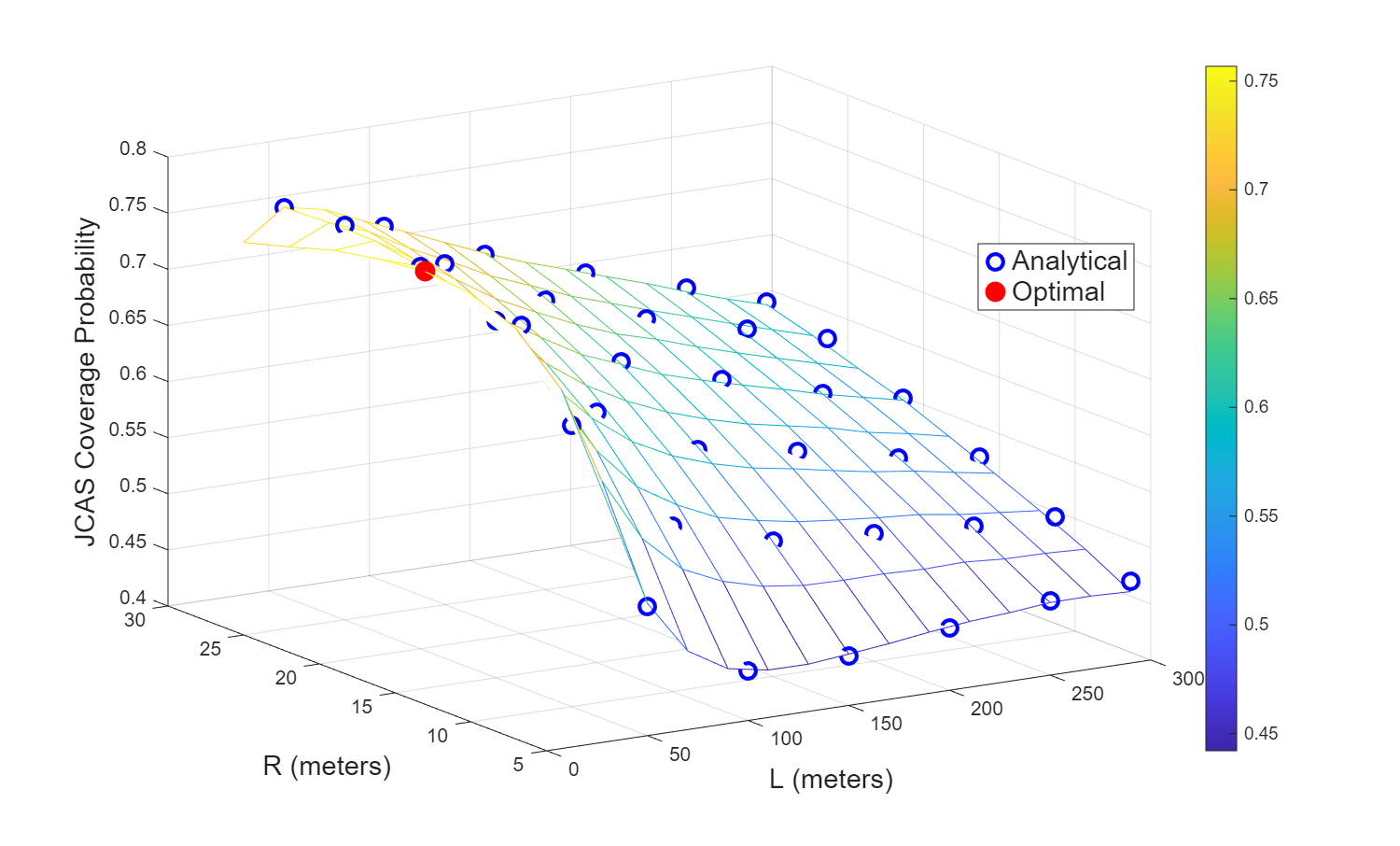}
       \caption{JCAS CP versus $L$ and $R$.}
    \label{fig:Fig4}
\end{figure}

Fig. \ref{fig:Fig4} presents the  analytical (denoted by markers) JCAS CP, obtained by Theorem 1, versus $L$ and $R$ for angular standard deviation equal to $4.5^{o}$. The first observation is that JCAS performance varies more strongly along the $L$ dimension than along the $R$ dimension.  This is because $L$ directly controls the axial spread of the UAV-UEs and, therefore, the pathloss function. In contrast, $R$ only controls the spread around the height of the corridor. Therefore, increasing $L$ has a stronger impact on the desired-link pathloss and LoS probability. Moreover, there is an optimal pair value $(L,R)$ to maximize the JCAS performance, which is observed for the smallest value of the length of the corridor.

\section{Conclusion}
A novel stochastic geometry framework based on 3D BPP was proposed to investigate the JCAS performance of UAV-assisted networks, with UAV-UEs operating in swarms in UAV corridor-enabled networks. Assuming that a terrestrial BS equipped with a 3D 3GPP antenna exploits JCAS capabilities, the performance of the  JCAS CP was  investigated under 3D beam misalignment. The misalignment error was modeled for the first time using the multivariate normal distribution. Among several key system-level insights, it was shown that: i) for the higher deployment heights, the JCAS performance decreases with an increase of corridor length until eventually stabilizes and ii) narrower beams reduce interference more effectively, but they are  also more sensitive to pointing errors.

\appendices
\section{Proof of Lemma 2}
The Laplace transform of $I_s$ conditioned on $\mathbf{w}_0$ and $\mathbf{\epsilon}$ is given by \eqref{LTproof}, shown at the bottom of the next page,
\begin{figure*}[!b]
\hrulefill
\begin{equation} \label{LTproof} 
\begin{split}
&\mathcal{L} _{I_s}(s|\mathbf{w}_0,\mathbf{\epsilon})=\mathbb{E}[e^{-s I_s}]=\mathbb{E}\Big[e^{-s \sum_{i \in \Phi^{!}}  g_{{\rm{mis}}}^{\boldsymbol{\epsilon}}(\theta_i,\phi_i)^2   h_{i}^s \sigma_i  l(\mathbf{w}_i)^2}\Big|\mathbf{w}_0,\mathbf{\epsilon}\Big] \\
&\myeqa \mathbb{E}_{\Phi^{!},\sigma_i}\Bigg[\prod_{i \in \Phi^{!}}\Bigg[ \Big(1+\frac{s\, g_{{\rm{mis}}}^{\boldsymbol{\epsilon}}(\theta_i,\phi_i)^2 \sigma_i \,   )}{m_t l^{-2}(\mathbf{w}_i)} \Big)^{-m_t} \Big|\mathbf{w}_0,\mathbf{\epsilon} \Bigg] \Bigg] \myeqb \Bigg[\mathbb{E}_{\Phi^{!}}\Bigg[ \int_{0}^{\infty}\Big(1+\frac{s\,  \sigma_i  g_{{\rm{mis}}}^{\boldsymbol{\epsilon}}(\mathbf{w}_i|\mathbf{w}_0)^2     )  }{m_t l^{-2}(\mathbf{w}_i)} \Big)^{-m_t} f_{\sigma_i}(\sigma_i) {\rm{d}}  \sigma_i \Big|\mathbf{\epsilon} \Bigg] \Bigg]^{N-1} \\
&\myeqc \Bigg[\mathbb{E}_{\mathbf{w}_i}\Bigg[ \int_{0}^{\infty}\Big(1+\frac{s\,  \sigma_i  g_{{\rm{mis}}}^{\boldsymbol{\epsilon}}(\mathbf{w}_i|\mathbf{w}_0)^2      }{m_L l^{-2}(\mathbf{w}_i |LoS )} \Big)^{-m_L } f_{\sigma_i}(\sigma_i) {\rm{d}}  \sigma_i \Big|\mathbf{\epsilon}+ \int_{0}^{\infty}\Big(1+\frac{s\,  \sigma_i  g_{{\rm{mis}}}^{\boldsymbol{\epsilon}}(\mathbf{w}_i|\mathbf{w}_0)^2      }{m_N l^{-2}(\mathbf{w}_i |NLoS )} \Big)^{-m_N} f_{\sigma_i}(\sigma_i) {\rm{d}}  \sigma_i \Big|\mathbf{\epsilon}  \Bigg]^{N-1} .
\end{split}
\end{equation}
\end{figure*}
where (a) follows from the  moment generating function of $h_{i}^s$ and the independent and identically distributed nature of all random variables, (b) follows after averaging over all RCS values of the interfering UAV-UEs and (c) follows after conditioning over the LoS/NLoS states of the UAV-UEs' spatial location. Finally, Lemma 2 is obtained after deconditioning over the LoS/NLoS states and averaging with the PDF of the 3D spatial location of interfering UAV-UEs by exploiting Lemma 1.

\section{Proof of Theorem 1}
Conditioning on the LoS/NLoS state of the tracked UAV-UE, the JCAS CP is decomposed as
\begin{align} \label{eq1}
&\mathcal{P}_{cs}(\theta_c,\theta_s)\\
&=\mathbb{E}_{\mathbf{w}_0,\boldsymbol{\epsilon}}
\Big[
\mathbb{P}\big(
{\rm SIR}_c>\theta_c,
{\rm SIR}_s>\theta_s
\mid
\mathbf{w}_0,\boldsymbol{\epsilon}
\big)
\Big] \nonumber\\
&=
\mathbb{E}_{\mathbf{w}_0,\boldsymbol{\epsilon}}
\Big[
P_L(\mathbf{w}_0)
\mathbb{P}\big(
{\rm SIR}_c>\theta_c,
{\rm SIR}_s>\theta_s
\mid
\mathbf{w}_0,\boldsymbol{\epsilon},L
\big)
\nonumber\\ \nonumber
&\qquad\qquad+
P_N(\mathbf{w}_0)
\mathbb{P}\big(
{\rm SIR}_c>\theta_c,
{\rm SIR}_s>\theta_s
\mid
\mathbf{w}_0,\boldsymbol{\epsilon},N
\big)
\Big]. \end{align} Since sensing and communication aggregate interference terms are independent, the conditional joint event factors are given by
\begin{align} \label{eq2}
&\mathbb{P}\big(
{\rm SIR}_c>\theta_c,
{\rm SIR}_s>\theta_s
\mid
\mathbf{w}_0,\boldsymbol{\epsilon},t
\big)
\nonumber\\
&\qquad =
\mathbb{P}\big(
{\rm SIR}_s>\theta_s
\mid
\mathbf{w}_0,\boldsymbol{\epsilon},t
\big)
\mathbb{P}\big(
{\rm SIR}_c>\theta_c
\mid
\mathbf{w}_0,\boldsymbol{\epsilon},t
\big)
\end{align} The two factors in \eqref{eq2} correspond to the expressions derived in Proposition 1 and Proposition 2, respectively. Therefore, after inserting Proposition 1 and Proposition 2 in \eqref{eq1} and averaging over $\mathbf{w}_0,\boldsymbol{\epsilon}$, yields Theorem 1. 

\section*{Acknowledgment}
This work was co-funded by the University of Piraeus Research Center (UPRC) and the EU HORIZON programme under iSEE-6G GA No. 101139291.

\bibliographystyle{IEEEtran}
\bibliography{IEEEabrv,references}
\balance

\end{document}